\documentclass[twocolumn,showpacs,preprintnumbers,amsmath,amssymb,prb,superscriptaddress]{revtex4}
\usepackage{graphicx}
\usepackage{epsf}
\usepackage{bm}
\usepackage{hyperref}
\usepackage{color}

\begin{document}

\title{In-plane magnetic field effect on hole cyclotron
mass and $g_z$ factor in high-mobility SiGe/Ge/SiGe structures}

\author{I.L.~Drichko}
\author{V.A.~Malysh}
\author{I.Yu.~Smirnov}
\author{L.E.~Golub}
\author{S.A.~Tarasenko}
\affiliation{A. F. Ioffe Physico-Technical Institute of Russian
Academy of Sciences, 194021 St.Petersburg, Russia}
\author{A.V.~Suslov}
\affiliation{National High Magnetic Field Laboratory, Tallahassee,
FL 32310, USA}
\author{O.A.~Mironov}
\affiliation{Department of Physics, University of Warwick, Coventry,
CV4 7AL, United Kingdom}
\affiliation{International Laboratory of
High Magnetic Fields and Low Temperatures, 53-421 Wroclaw, Poland}
\author{M.~Kummer}
\author{H.~von~K\"{a}nel}
\affiliation{Laboratorium f$\ddot{u}$r Festk\"{o}rperphysik ETH
Z\"{u}rich, CH-8093 Z\"{u}rich Switzerland}


\begin{abstract}
The high-frequency (ac) conductivity
of a high quality modulation doped GeSi/Ge/GeSi single quantum well
structure with hole density $p$=6$\times$10$^{11}$cm$^{-2}$ was
measured by the surface acoustic wave (SAW) technique at frequencies
of 30 and 85~MHz and magnetic fields $B$ of up to 18 T in the
temperature range of 0.3 -- 5.8 K. The acoustic effects were also
measured as a function of the tilt angle of the magnetic field with
respect to the normal of the two-dimensional channel at $T$=0.3 K.
It is shown, that at the minima of the conductivity oscillations,
holes are localized on the Fermi level, and that there is a
temperature domain in which the high-frequency conductivity in the
bulk of the quantum well is of the activation nature. The analysis
of the temperature dependence of the conductivity at odd filling
factors enables us to determine the effective $g_z$ factor. It is
shown that the in-plane component of the magnetic field leads to an
increase of the cyclotron mass and to a reduction of the $g_z$
factor. We developed a microscopic theory of these effects for the
heavy-hole states of the complex valence band in quantum wells which
describes well the experimental findings.
\end{abstract}


\maketitle

\section{Introduction}\label{Introduction}

Modulation doped SiGe/Ge/SiGe structures with a two-dimensional (2D)
hole gas are attractive systems  for both fundamental and applied
studies since they are compatible with silicon-based technology and,
at the same time, have a record high hole mobility among all group
IV semiconductors. As compared to silicon-based
metal-oxide-semiconductor field-effect-transistor structures, they
are also characterized by a strong spin-orbit coupling and a large
and strongly anisotropic $g$-factor, which is of interest for the
study of spin-related phenomena. However, the details of the band
structure and the quantum transport in $p$-SiGe/Ge/SiGe systems have
not yet been sufficiently explored. It is known, that due to the
lattice constant mismatch the 2D hole channel is located in strained
Ge so that, the ground subband is formed by heavy-hole (hh) states
while the light-hole (lh) subband is split off by a hundred
meV.~\cite{kanel} The splitting should suppress the hh-lh mixing and
lead to a strong anisotropy of the hh $g$-factor tensor with
vanishingly small in-plane component $g_{\parallel}$.~\cite{IP_book}
One can therefore expect that the transport properties of the hole
channel in SiGe/Ge/SiGe are determined by the normal component of
the magnetic field only,~\cite{Tarasenko02} as they are in a
strained $p$-type channel of A$^{\text {III}}$B$^{\text V}$
semiconductors.~\cite{Martin,Dorozhkin,Lin} Such a behavior has been
 observed in $p$-Ge/SiGe multilayer structures by
studying the resistivity oscillations in tilted magnetic fields up
to a tilt angle of 60$^\circ$.~\cite{Gorodilov,Arapov1} What
concerns us about the absolute value of the hh out-of-plane
$g$-factor, $|g_z|$, is that the data available in literature vary
from 20.4 for bulk Ge,~\cite{Arapov3} to 16.5 for strained
Ge,~\cite{Nenashev} to 14.2 and 5.8 for $p$-Ge/GeSi
multilayers,~\cite{Arapov3,Chernenko} and to 1 for a single quantum
well.~\cite{Moriya2013} These drastically different values may be
the result of many-body effects and spectrum
nonparabolicity~\cite{IP_book} which are also responsible for the
observed dependence of the in-plane effective mass on the hole
density.\cite{Irisawa,Rossner}

In this paper, we report a comprehensive study of the high-frequency
conductivity of the high-mobility 2D hole gas embedded in a
SiGi/Ge/SiGe structure in tilted magnetic fields. The measurements
are carried out by means of the contactless surface acoustic wave
technique. It probes the ``bulk'' electric properties of the
two-dimensional system and provides information on the 2D hole
energy spectrum unaffected by chiral edges which play a key role in
conventional four-probe measurements of the quantum Hall effect. The
experimental data allow us to determine the hole cyclotron mass from
the temperature dependence of Shubnikov-de~Haas (SdH) oscillations
and the effective $g_z$ factor for our samples. We find that both
the cyclotron mass and the $g_z$ factor can be tuned by applying an
in-plane magnetic field. Raising the in-plane field component
$B_{\parallel}$ leads to an increase of the cyclotron mass and
decrease of the $g_z$ factor. We have developed a microscopic theory
of these effects for the complex valence band of germanium and have
shown that the theory describes well the experimental data.

\section{Sample and Method}\label{Sample_Method}

\begin{figure}[ht]
\centerline{
\includegraphics[width=9cm]{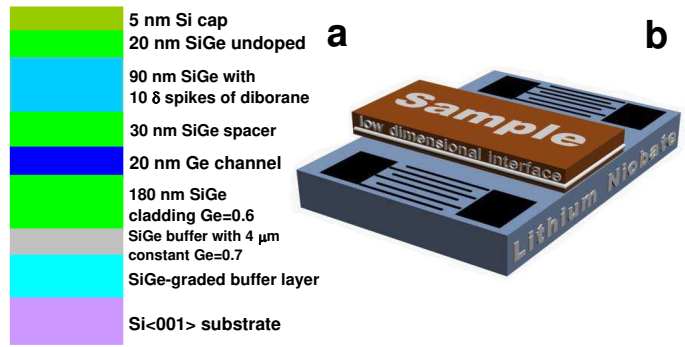}}
\caption{(Color online) (a) Cross-section of the studied sample
and (b) sketch of the acoustic experimental setup.
\label{Sample}}
\end{figure}

The experiments were carried out on a $p$-type SiGe/Ge/SiGe
heterostructure with a single hole channel (sample K6016).
\cite{kanel,K6016our} The layer structure of the sample is
illustrated in Fig.~\ref{Sample}a. The two-dimensional hole gas has
a density of $p \approx 6 \times$10$^{11}$cm$^{-2}$ and mobility
$\mu \approx 6 \times 10^4$~cm$^2$/(V$\cdot$s) at 4.2~K. The
structure was grown by low-energy plasma-enhanced chemical vapor
deposition (LEPECVD) on a Si(001) substrate by making use of the
large dynamic range of growth rates. \cite{lepecvd} The buffer,
graded at the rate of about 10$\% / \mu$m to the final Ge content of
70$\%$, and the 4-$\mu$m-thick Ge$_{0.7}$Si$_{0.3}$ layer were grown
at the high rate of $5\div10$ nm/s gradually lowering the substrate
temperature $T_s$ from 720$^\text{o}$C to 450$^\text{o}$C. The
active part, consisting of a 20-nm-thick pure Ge layer sandwiched
between cladding layers with a Ge content of about 60$\%$, and the
Si cap were grown at the low rate of about 0.3~nm/s at $T_s$ =
450$^\text{o}$C. Modulation doping was achieved by introducing
dilute diborane pulses into the cladding layer at a distance of
about 30 nm above the channel. In the structure grown on a relaxed
Si$_{0.3}$Ge$_{0.7}$ buffer layer, the Ge channel is compressively
strained in the interface plane due to the lattice mismatch of about
$1.2$\%, which leads to a splitting between the hh and lh subbands.
Following Ref.~\onlinecite{Walle82}, we estimate the hh-lh splitting
$\Delta \approx 100$~meV for our sample.

The properties of the 2D hole gas are studied by a
contactless acoustoelectric method. The technique was first employed
in Ref.~\onlinecite{Wixforth} for GaAs/Al$_x$Ga$_{1-x}$As
heterostructures. The experimental setup is illustrated in
Fig.~\ref{Sample}(b). A surface acoustic wave (SAW) is excited on
the surface of a piezoelectric LiNbO$_3$ platelet by an
inter-digital transducer. The SAW propagating along the lithium
niobate surface induces a high-frequency electric field which
penetrates into the hole channel located in the SiGe/Ge/SiGe
structure slightly pressed to the piezoelectric platelet by means of
springs. The field produces an ac electrical current in the channel.
As the result of the interaction of the SAW electric field
with holes, the wave attenuates and its velocity is modified,
governed by the high-frequency conductivity $\sigma^{ac}$. This
``sandwich-like'' experimental configuration enables contactless
acoustoelectric experiments on non-piezoelectric 2D systems, such as
SiGe/Ge/SiGe. The measurements were done at SAW frequencies of 30
and 85 MHz, in external magnetic fields $B$ of up to 18~T, and in
the temperature range of 0.3 -- 5.8 K. The samples were mounted on a
one-axis rotator, which enabled us to change the angle $\Theta$
between the quantum well (QW) normal and the magnetic field.

\section{Experimental results}

\begin{figure}[ht]
\centerline{
\includegraphics[width=9cm,clip=]{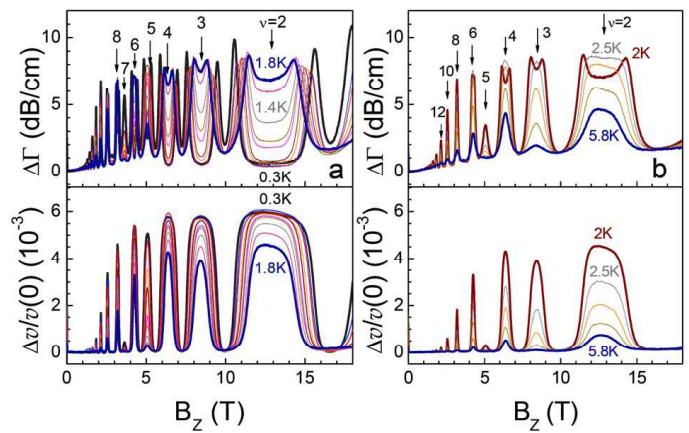}}
\caption{(Color online) Magnetic field dependences of $\Delta \Gamma$ and
$\Delta v/v(0)$ measured in a field $B \parallel z$ up to 18~T at the SAW frequency
$f=30$~MHz, and in the temperature ranges of (a) 0.3--1.8~K and
(b) 2--5.8~K. Arrows denote the positions of integer filling factors.
 \label{fig:GV1}}
\end{figure}

Figure~\ref{fig:GV1} shows the dependencies of the SAW attenuation
change $\Delta\Gamma \equiv \Gamma(B) - \Gamma(0)$ and normalized
change of the SAW velocity $\Delta v/v(0) \equiv [v(B) - v(0)]/v(0)$
on the magnetic
 field $B$ applied along the QW normal $z$. The dependencies are
 presented for different
temperatures. All curves contain pronounced oscillations, which are
caused by the formation of Landau levels in the two-dimensional hole
gas and microscopically are similar to the oscillations of the
magnetoconductivity in the Shubnikov-de Haas and quantum Hall
effects. We note that $\Gamma(0)$ is negligibly small compared to
$\Gamma (B)$ due to very high electric conductivity of the hole gas
at zero magnetic field, $\sigma(0) \approx 6 \times 10^{-3} \:
\Omega^{-1}$.

\begin{figure}[h]
\centerline{
\includegraphics[width=9cm,clip=]{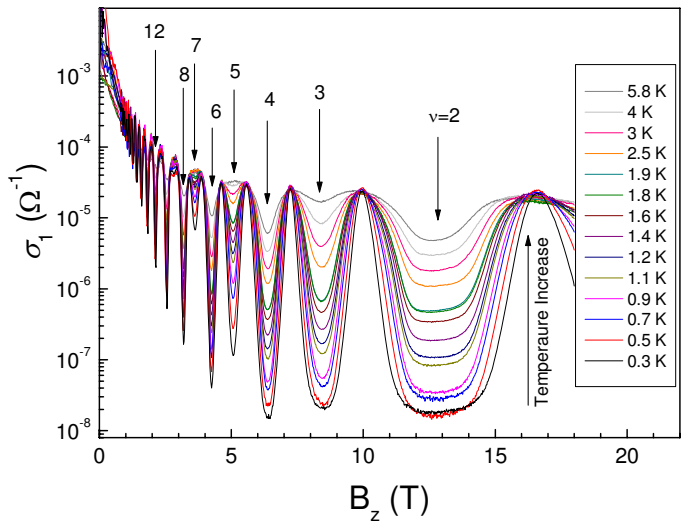}}
\caption{(Color online) Dependence of $\sigma_{1}$ on the magnetic field at different temperatures for
$B \parallel z$ and for a SAW frequency $f = 30$~MHz. The positions of integer filling factors are marked by arrows.}
\label{S1}
\end{figure}

From the experimentally measured values of the SAW absorption and
the relative change of the SAW velocity, one can calculate the real
$\sigma_{1}$ and imaginary $\sigma_{2}$ components of the
high-frequency conductivity of the hole channel by using Eqs.~(1)
and~(2) of Ref.~\onlinecite{K6016our}. Below, we focus on the real
part of the ac conductivity only since it enables us to determine
the parameters of the energy spectrum. The corresponding dependence
of $\sigma_{1}$ on the magnetic field for different temperatures is
presented in Fig.~\ref{S1}.  The magnetic field dependence of the
conductivity contains the Shubnikov-de Haas oscillations evolving
into the integer quantum Hall effect at strong fields. The positions
of the even and odd filling factors $\nu$ corresponding to the
orbital and spin splitting of the Landau levels, respectively, are
shown by vertical arrows.

\begin{figure}[ht]
\centerline{
\includegraphics[width=8.9cm]{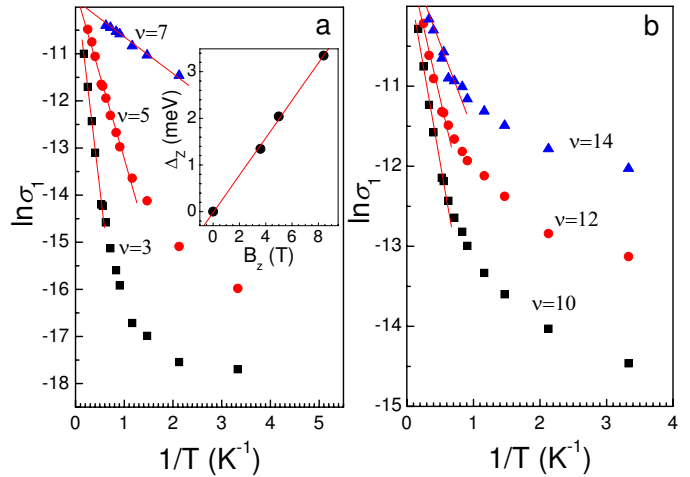}
}
\caption{(Color online) Dependence of $\ln \sigma_1$ on $1/T$ for
(a)~odd and (b)~even filling factors. Lines are the result of linear fitting.
 Inset shows the obtained Zeeman splitting
vs perpendicular magnetic field for odd filling factors.
 \label{FigDEspin}}
\end{figure}

Comparison of the temperature dependence of the conductivity at odd
and even filling factors allows us to determine the absolute value
of the $g_z$ factor. The procedure is the following. For odd filling
factors, we found a temperature range (0.5-5.8~K) where the
conductivity in the oscillation minima is of activation nature and
described by the Arrhenius law
\begin{equation}\label{Arrhenius}
\sigma_1^{\text{odd}} \propto \exp\left( - \frac{\Delta_{\rm odd}}{2k_B T} \right) \:.
\end{equation}
Here $\Delta_{\rm odd}$ is the activation energy and $T$ is the temperature. Thus,
the slope of the linear dependence of $\ln \sigma_1$ on $1/T$ yields the activation energy.
The corresponding Arrhenius plots for the filling factors $\nu$=3,
5, and 7 together with linear fits are presented in
Fig.~\ref{FigDEspin}a.
The activation energy is given by $\Delta_{\rm odd} = \Delta_Z - \Gamma_B$, where
$\Delta_Z = |g_z| \mu_0 B_z$ is the Zeeman splitting, $\mu_0$ is the Bohr magneton,
and $\Gamma_B$ is the Landau level broadening. The latter also depends on the magnetic field,
the calculation in the self-consistent Bohr approximation yields $\Gamma_B = C \sqrt{B_z}$ with $C$
being the field-independent parameter.~\cite{Ando,Coleridge2003}

For even filling factors, the activation conductivity
is also described by the Arrhenius law
\begin{equation}\label{Arrhenius_even}
\sigma_1^{\text{even}} \propto \exp\left( - \frac{\Delta_{\rm even}}{2k_B T} \right) \:,
\end{equation}
where $\Delta_{\rm even} = \hbar\omega_c - \Delta_Z - \Gamma_B$,
$\hbar\omega_c = \hbar e B_z/(m_c c)$ is the energy spacing between
the orbital Landau levels and $m_c$ is the cyclotron mass. The
cyclotron mass for our Ge/SiGe structure $m_c \approx 0.1 m_0$ is
known with high accuracy from analysis of the Shubnikov-de Haas
oscillations.~\cite{K6016our} The Arrhenius plots for the even
filling factors $\nu$=10, 12, and 14 together with linear fits are
presented in Fig.~\ref{FigDEspin}b.

The best fit of the activation conductivity for odd and even filling
factors by Eqs.~\eqref{Arrhenius} and~\eqref{Arrhenius_even} yields
$|g_z| = 6.7 \pm 0.3$ and $C = 0.64 \pm 0.06$~meV T$^{-1/2}$. The
extracted dependence of the Zeeman splitting $\Delta_Z$ on the
magnetic field is shown in the inset in Fig.~\ref{FigDEspin}a. Note,
that the theoretical estimation $C=\sqrt{2 e \hbar^2/(\pi m_c c
\tau_q)}$ (see Ref.~\onlinecite{Coleridge2003}) yields the very
close value $C \approx 0.69$~meV~T$^{-1/2}$ for the quantum
relaxation time $\tau_q \approx 1$~ps calculated for our sample from
the Shubnikov-de Haas oscillations at the temperature
1.7~K.~\cite{K6016our}
%
The obtained value of the out-of-plane $g$-factor differs from that
of the heavy holes in bulk Ge, $|g_{{\rm bulk}}| = |6 {\cal K} |
\approx 20.4$. We attribute the difference to renormalization of the
energy spectrum in Ge quantum wells due to size quantization,
strain, and interaction effects. The dispersion of heavy holes in
quantum wells is typically non-parabolic so that the in-plane
effective mass and $g$-factor depend on the Fermi energy. We also
note that in the above analysis we neglected possible oscillations
of the hole $g$-factor in the magnetic field due to exchange
interaction. While the exchange contribution to the g-factor may be
important for 2D electron systems, experimental results and
theoretical analysis reveal that it is suppressed in GaAs-based hole
systems.~\cite{Winkler2005,Chiu2011,Kernreiter2013} The problem of
the exchange interaction in strained Ge-based systems requires
further study and is out of the scope of this paper.

\begin{figure}[h]
\centerline{
\includegraphics[width=8.5cm]{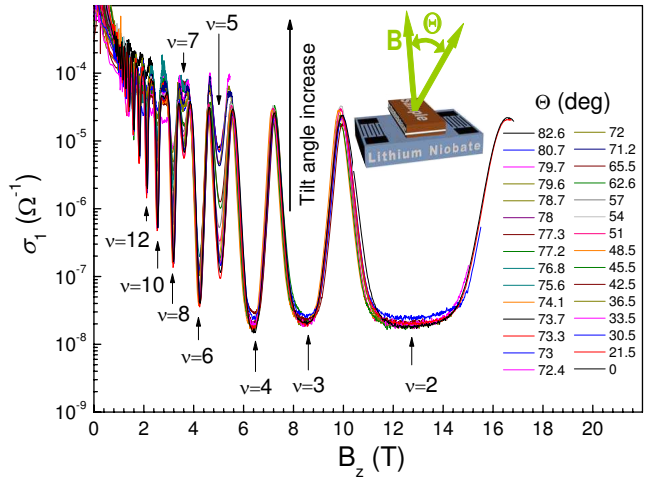}
} \caption{(Color online) Dependence of $\sigma_1$ on the normal
component $B_z$ of the magnetic field for different tilt angles
$\Theta$=(0$\div$82)$^{\text o}$; $f$=30 MHz, $T$=0.3 K.
 \label{S1tilt}}
\end{figure}

\begin{figure}[h]
\centerline{
\includegraphics[width=8.9cm]{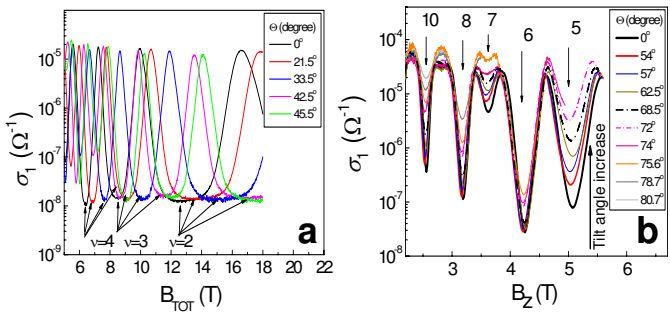}
} \caption{(Color online) (a) Dependence of $\sigma_1$ on the total magnetic
field $B_{{\rm TOT}}$ for different tilt angles; (b) $\sigma_1$ vs $B_z$ for the tilt angles
$\Theta$=0$^{\text o}$ - 80$^{\text o}$; $f$=30~MHz, and
$T$=0.3~K. Arrows denote the positions of integer filling factors.
 \label{S1tilt_total}}
\end{figure}

To study the $g$-factor anisotropy, the SAW absorption and velocity
change were measured in tilted magnetic fields of magnitude
$B_{\text {TOT}}$. Figure~\ref{S1tilt} shows the obtained dependence
of the real part of the conductivity $\sigma_1$ on the normal
component of magnetic field $B_{z}=B_{\text {TOT}} \cos\Theta$ for
various tilt angles $\Theta$. One can see that the positions of the
conductivity minima are determined by the normal component of the
magnetic field, which is in agreement with the 2D character of the
hole states. Such behavior comes from the significant strain in
$p$-SiGe/Ge/SiGe structures resulting in the splitting between the
hh and lh subbands of about 100~meV which exceeds the Fermi energy
of 14~meV. Therefore, the in-plane component of the hh $g$-factor
vanishes and the Zeeman splitting is determined by
$B_z$.\cite{footnote1} Similar dependencies are also presented in
Fig.~\ref{S1tilt_total}a where the conductivity oscillations
corresponding to small filling factors $\nu =2$, 3, and 4 are shown
as a function of the total magnetic field. With increasing the tilt
angle, the positions of the oscillation minima are shifted towards
higher magnetic fields while the oscillation amplitudes remain
almost the same. Note, however, that study of conductivity at small
filling factors requires high $B_z$ and therefore for $\nu$=4 we
were limited by 60$^\circ$ tilt angle.

Figure~\ref{S1tilt_total}b shows the magnetoconductivity
oscillations as a function of $B_z$ at various angles $\Theta$ for
large filling factors $\nu \geq 5$, where a  strong in-plane field
$B_{\parallel}$ can be applied to the sample. Surprisingly, we
observe an effect of $B_{\parallel}$ on the oscillations: The
conductivity in the minima increases with increasing tilt angle at
large $\Theta$ and the oscillation amplitude decreases. We emphasize
that such a behavior is observed for both even and odd filling
factors, see Fig.~\ref{S1tilt_total}b. Therefore, it cannot be
explained by changes in relative positions of the Landau levels
since, in that case, the amplitudes of conductivity oscillations
corresponding to even and odd $\nu$ would change in antiphase. We
attribute the observed features to the effect of the in-plane
magnetic field on the hole cyclotron mass, $g_z$ factor, and Landau
level broadening in the complex valence band of germanium.

\begin{figure}[h]
\centerline{
\includegraphics[width=6.7cm]{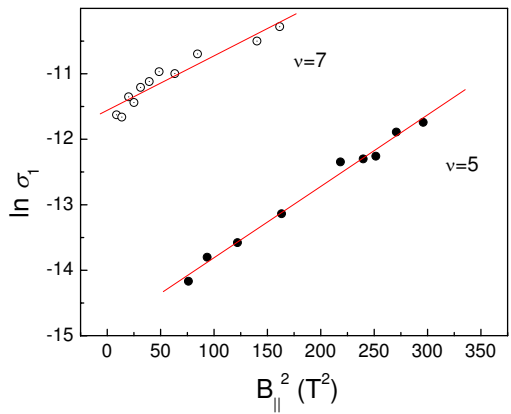}
} \caption {
Dependence of $\ln \sigma_1$
on $B_\parallel^2$ for the filling factors $\nu$=5 and~7.
\label{Btan}}
\end{figure}

According to the Arrhenius law Eq.~(\ref{Arrhenius}), the decrease
of the oscillation amplitudes at odd filling factors indicates a
decrease of the activation energy $\Delta_{\rm odd}$. We suggest
that the decrease is caused by the reduction of the absolute value
of the hole $g$-factor $|g_z|$ and increase of the Landau level
broadening in the in-plane magnetic field. To second order in
$B_{\parallel}$, the dependence of $g_z$ and $C$ on the magnetic
field is given by
\begin{eqnarray}\label{g_phen}
g_z(B_{\parallel}) &=& g_z(0) + \alpha_s B_{\parallel}^2 \:, \\
C(B_{\parallel}) &=& C(0) + \beta B_{\parallel}^2 \:, \nonumber
\end{eqnarray}
where $g_z(0)$ and $C(0)$ are the $g$-factor and the parameter of
the Landau level broadening at zero in-plane field, which are
calculated above,  $\alpha_s$ and $\beta$ are parameters. Below, see
part IV, we present the microscopic theory of the Zeeman splitting
of heavy-hole states in QWs and show that exactly such a dependence
of the $g_z$-factor on the in-plane field follows from the theory.
Equations~\eqref{Arrhenius} and~(\ref{g_phen}) yield
\begin{equation}\label{sigma_theta_g}
\ln \sigma_1^{\text{odd}}(B_{\parallel}) =
\ln \sigma_1^{\text{odd}}(0)
+ \frac{\alpha_s \mu_0 B_z + \beta \sqrt{B_z}}{2 k_B T} B_{\parallel}^2 \:,
\end{equation}
where we take into account that $g_z < 0$ for heavy holes in Ge, and
$\sigma_1^{\text{odd}}(0)$ is the conductivity in perpendicular
magnetic field for given odd filling factor (an in-plane field
independent term). To determine the field corrections to both the
$g_z$ factor and Landau level broadening we plot in Fig.~\ref{Btan}
the dependence of $\ln\sigma_1(B_{\parallel})$ on $B_{\parallel}^2$
for $\nu=5$ and~$7$ in the range of activation behavior of
conductivity. The dependencies are linear, as expected, and yield
$\alpha_s^{({\rm exp})} \approx 1.4 \times 10^{-3}$~T$^{-2}$,
$\beta^{({\rm exp})} \approx 8\times 10^{-5}$~meV~T$^{-5/2}$. For
the magnetic field $B_z=5.04$~T corresponding to $\nu=5$ the effect
of $B_\parallel$ on the Zeeman splitting is more than twice as large
as the effect of $B_\parallel$ on the level broadening.

The oscillations corresponding to even filling factors are also
damped out with increasing the tilt angle $\Theta$, see
Fig.~\ref{S1tilt_total}b. This indicates a decrease of the
activation energy $\Delta_{\rm even}$ which can be attributed to the
increase of the cyclotron mass $m_c$ and the Landau level broadening
by the in-plane component of the magnetic field. Similar behavior
was observed for $n$-type 2D systems in a number of papers and
ascribed to the increase of $m_c$.~\cite{Khrapai,Kukushkin,Zudov} To
second order in $B_{\parallel}$, the effect is phenomenologically
described by
\begin{equation}\label{m_phen}
\frac{m_0}{m_c(B_{\parallel})} = \frac{m_0}{m_c(0)} - \alpha_c B_{\parallel}^2 \:,
\end{equation}
where $m_0$ is the free electron mass and $\alpha_c$ is a parameter.
From the Arrhenius law at even
filling factors, Eq.~\eqref{Arrhenius_even}, we obtain
\begin{equation}\label{sigma_theta_m}
\ln \sigma_1^{\text{even}}(B_{\parallel})
= \ln \sigma_1^{\text{even}}(0) + \frac{(2\alpha_c - \alpha_s) \mu_0 B_z +
\beta \sqrt{B_z}}{2 k_B T}  B_{\parallel}^2\:,
\end{equation}
where $\sigma_1^{\text{even}}(0)$ is the conductivity in
perpendicular magnetic field for given even filling factor (the term
independent of $B_{\parallel}$).

\begin{figure}[h]
\centerline{
\includegraphics[width=7cm]{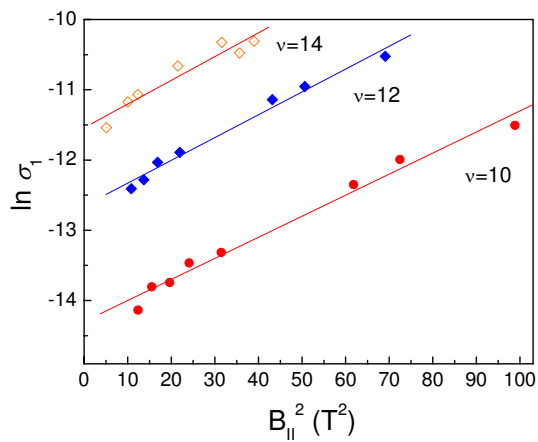}
}
\caption {(Color online) Dependence of $\ln \sigma_1$ on
$B_\parallel^2$ at the conductivity minima corresponding to the even filling factors $\nu$=10, 12, and 14.
\label{s1tan}}
\end{figure}

Figure~\ref{s1tan} shows the dependence of $\ln
\sigma_1(B_{\parallel})$ on $B_{\parallel}^2 $ measured at different
$B_z$ corresponding to the filling factors $\nu = 10$, 12, and 14.
In accordance with Eq.~(\ref{sigma_theta_m}), the dependencies are
linear. Fitting the experimental data by Eq.~\eqref{sigma_theta_m}
with the parameters $\alpha_s$ and $\beta$ obtained above  yields
$\alpha_c^{({\rm exp})} \approx 6 \times 10^{-3}$~T$^{-2}$.
Comparing $2\alpha_c^{({\rm exp})}$ with $\alpha_s^{({\rm exp})}$ we
conclude that the dominant contribution to the variation of
$\Delta_{\rm even}$ is due to change of the cyclotron mass.

Below we calculate the parameters $\alpha_s$ and $\alpha_c$
determining the corrections to the spin splitting and cyclotron
mass, respectively, for the heavy-hole subband in QWs and compare
them with the values obtained from our experiment.

\section{Theory}\label{sec:theory}

We describe the effect of the in-plane magnetic field on the
cyclotron mass and $g_z$ factor in the framework of the Luttinger
model. In the axial approximation, the effective Hamiltonian of
holes in Ge quantum wells in an external magnetic field has the form
\begin{equation}
{\cal H} = H_0 + U(z) + H_Z + V \:,
\end{equation}
where $H_0$ is the Luttinger Hamiltonian for zero in-plane momentum,
\begin{equation}
    H_0 = {1 \over 2 m_0}\left[\left(\gamma_1 + {5\over 2}\gamma\right) I - 2 \gamma J_z^2 \right] p_z^2 \:,
\end{equation}
$\gamma_1$ and $\gamma$ are the Luttinger parameters, $J_i$ ($i=x,y,z$) are the $4\times4$ matrices of the angular momentum $3/2$, $I$ is the identity matrix, $p_z = - {\rm i} \hbar \partial /\partial z$, $U(z)$ is the diagonal matrix of confinement potentials which are different for the heavy-hole and light-hole subbands due to strain, $H_Z$ is the Zeeman Hamiltonian,
\begin{equation}
H_Z =  - 2 {\cal K} \mu_0 (\bm{J} \cdot \bm{B}) \:,
\end{equation}
${\cal K}$ is the parameter of the Zeeman splitting of hole states at the $\Gamma$ point of the Brillouin zone in bulk material,
$V$ is the contribution to the Luttinger Hamiltonian accounting for the in-plane momentum and magnetic field,

\begin{align}
    V = & \frac{1}{2m_0} \left(\gamma_1 + {5\over 2}\gamma \right) I \left( P_x^2+P_y^2 \right) \\
    & - {\gamma \over m_0} \left(J_x^2 P_x^2 + J_y^2 P_y^2 + 2 \{J_x J_y\}\{P_x P_y\} \right), \nonumber
\end{align}
$\bm P = -{\rm i} \hbar \nabla - (e/c)\bm{A}$, $e>0$ is the hole
charge, $\bm{A}$ is the vector potential of the magnetic field
$\bm{B}$, and the braces denote the symmetrized product $\{CD\} =
(CD+DC)/2$. Other possible contributions to the effective
Hamiltonian beyond the Luttinger model, e.g., $\propto B^2$ or
$\propto B^3$, do not seem to give a substantial contribution to the
cyclotron mass and $g_z$ factor in moderate in-plane magnetic
fields.

We calculate the hole energy spectrum in the symmetric
heterostructure subjected to the magnetic field in two steps.
First, we solve the Schr\"{o}dinger equation for the case of
$H_Z=V=0$
and find the envelope functions $\varphi_{hn}(z)$ and $\varphi_{lm}(z)$ and energies $E_{hn}$ and $E_{lm}$ of the heavy-hole and light-hole states,
respectively, where $n,m=1,2,\ldots$ are the subband indices.
Each state is two-fold degenerate. Then we use perturbation theory
and calculate the in-plane effective mass for the $h1$ subband,
which becomes anisotropic in the presence of the in-plane magnetic
field, as well as the Zeeman splitting at the subband bottom. We
assume that the magnetic field $\bm{B}$ is oriented in the $(yz)$
plane and choose the vector potential in the form $\bm{A}=(z B_y, x
B_z,0)$. The corrections to the in-plane mass and the Zeeman
splitting are proportional to $B_y^2$ and emerge in the second,
third, and fourth orders of the perturbation theory.
The knowledge of the in-plane effective masses and the Zeeman
splitting enables us to obtain the quasi-classical structure of the
Landau levels.

The calculation shows that the energy spectrum in the subband $h1$ has the form
\begin{equation}
E_{h1,\pm}^{(N)} = \frac{\hbar e B_z (N+1/2)}{c\, m_c(B_{y})} \pm \frac{\mu_0 \, g_{z}(B_{y}) B_z}{2} \:,
\end{equation}
where $m_c(B_y)=\sqrt{m_x(B_y) m_y(B_y)}$ is the cyclotron mass, $m_x(B_y)$ and $m_y(B_y)$ are the in-plane effective masses in the directions perpendicular and parallel to the in-plane component of the magnetic field, respectively, and
$N$ is an integer number.
The in-plane component of the
$g$-factor tensor vanishes at $B=0$ in the uniaxial
approximation~\cite{Marie98} and therefore gives only higher order
corrections ($\propto B^5$) to the Zeeman splitting for tilted
magnetic fields.

The $g_z$ factor determining the spin splitting of Landau levels is
given by Eq.~\eqref{g_phen} where the value in the perpendicular
field $g_z(0)$ has the form~\cite{Wimbauer94}
%
\begin{equation}
g_z(0) = - 6 {\cal K} + {12\gamma^2 \over m_0^2} \sum_n {| p^z_{h1,ln}|^2 \over E_{ln}-E_{h1}},
\end{equation}
$p^z_{h n, l m} = \langle h n| p_z | l m \rangle$ are the matrix elements of the momentum operator, and the coefficient $\alpha_s$ is given by
\begin{widetext}
    \begin{multline}
\label{alpha_s}
    \alpha_s={12\gamma^2 e^2 \over m_0 c^2}  \sum_{n,m,k} \Biggl[
{z_{h1,ln}^2 \over E_{ln} - E_{h1}} + {\gamma_1 + \gamma\over 2m_0}
     {(z^2)_{h1,h1}| p^z_{h1,ln}|^2 \over (E_{ln}-E_{h1})^2}
        - {\gamma_1 - \gamma\over 2m_0} {2 p^z_{h1,ln}z_{ln,lm}\{p_z z\}_{lm,h1} + p^z_{h1,ln}(z^2)_{ln,lm}p^z_{lm,h1} \over (E_{ln}-E_{h1})(E_{lm}-E_{h1})} \\
        -{\gamma_1 + \gamma \over m_0}{ (z^2)_{h1,hn}p^z_{hn,lm}p^z_{lm,h1}
        + z_{h1,hn}\left(\{p_z z\}_{hn,lm}p^z_{lm,h1}-p^z_{hn,lm}\{p_z z\}_{lm,h1}\right)
        \over (E_{hn}-E_{h1})(E_{lm}-E_{h1})}    \\
    +
        {3\gamma^2  \over m_0^2}  {p^z_{h1,ln}\{p^z z\}_{ln,hm}\{p^z z\}_{hm,lk}p^z_{lk,h1}
    - \{p^z z\}_{h1,ln}p^z_{ln,hm}p^z_{hm,lk}\{p^z z\}_{lk,h1}
    + 2p^z_{h1,ln}p^z_{ln,hm}\{p^z z\}_{hm,lk}\{p^z z\}_{lk,h1}\over (E_{ln}-E_{h1})(E_{hm}-E_{h1})(E_{lk}-E_{h1})}
         \\
        - {3\gamma^2  \over m_0^2} {|p^z_{h1,ln}|^2 |\{p^z z\}_{h1,lm}|^2 + |p^z_{h1,lm}|^2 |\{p^z z\}_{h1,ln}|^2
    \over (E_{ln}-E_{h1})^2(E_{lm}-E_{h1})} \Biggr] \:.
\end{multline}
\end{widetext}
Here  we take into account that the matrix
elements $z_{\mu n,\mu' m} = \langle \mu n|z| \mu' m \rangle$ and
$z^2_{\mu n,\mu' m} = \langle \mu n|z^2|\mu' m \rangle$ are real
while the matrix elements $p^z_{\mu n,\mu' m}$ are purely imaginary
($\mu,\mu'=h,l$).
The major contribution to $\alpha_s$ comes typically from the terms containing only one light-hole energy $E_{ln}$ in the denominator.

\begin{figure}[b]
\centerline{
\includegraphics[width=0.85\linewidth]{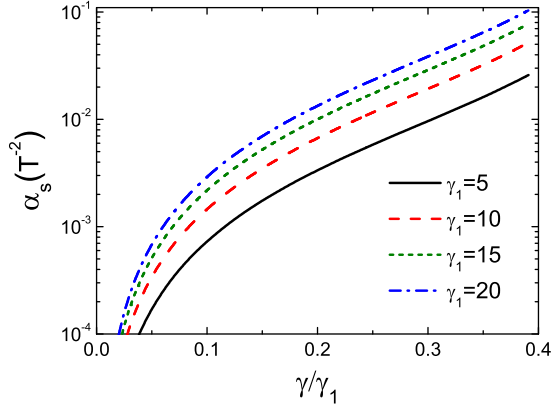}
} \caption {(Color online) Dependence of the coefficient $\alpha_s$
on the Luttinger parameters in strain-free QW of width
${a=200}$~\AA.} \label{fig_alpha_s}
\end{figure}
Figure~\ref{fig_alpha_s} shows the dependence of $\alpha_s$
responsible for the $g_z$ factor renormalization on the Luttinger
parameters calculated after Eq.~\eqref{alpha_s} for strain-free
rectangular QWs with infinitely high barriers. In this model, the
dependence of $\alpha_s$ on the QW width $a$ is simplified to
$\alpha_s \propto a^4$. The curves in Fig.~\ref{fig_alpha_s} are
plotted for $a=200$~\AA. It follows from the calculation that the
correction to the $g_z$ factor caused by the in-plane magnetic field
depends on the material parameters.

The in-plane masses are given by
\begin{equation}
    {m_0\over m_{x,y}(B_y)} = {m_0 \over m_\parallel} -(\alpha_c \pm \delta) B_y^2 \:,
\end{equation}
where $m_\parallel$ is the in-plane mass at zero magnetic
field,
%
\begin{equation}
    {1 \over m_\parallel} = {\gamma_1 + \gamma \over m_0} - {6\gamma^2 \over m_0^2} \sum_n {| p^z_{h1,ln}|^2 \over E_{ln}-E_{h1}} \:.
    \end{equation}
The parameter $\alpha_c$, which determines renormalization of the cyclotron mass in the in-plane field, and the parameter $\delta$ describing the mass anisotropy
have the form
\[
\alpha_c =  \alpha_s/2+\xi+\xi_1, \qquad \delta=\xi+\xi_2,
\]
where $\alpha_s$ is given by Eq.~\eqref{alpha_s},
\begin{widetext}
\begin{multline}
\label{xi}
    \xi =   {e^2 \over m_0 c^2} \sum_{m,n} \Biggl\{ {(\gamma_1 + \gamma)^2  z_{h1,hn}^2 \over E_{hn} - E_{h1}}
    -{36\gamma^4 \over m_0^2}
    {p^z_{h1,ln}\{p^z z\}_{ln,h1}p^z_{h1,lm}\{p^z z\}_{lm,h1}\over (E_{ln}-E_{h1})^2(E_{lm}-E_{h1})} \\
    + {6\gamma^2(\gamma_1+\gamma)\over m_0}  \left[
        {z_{h1,h1}p^z_{h1,ln}\{p_z z\}_{ln,h1}\over (E_{ln}-E_{h1})^2} -
        {z_{h1,hn}\left(\{p_z z\}_{hn,lm}p^z_{lm,h1}+p^z_{hn,lm}\{p_z z\}_{lm,h1}\right)
        \over (E_{hn}-E_{h1})(E_{lm}-E_{h1})}\right]
     \Biggr\}
                    \:,
    \end{multline}
\begin{equation}
\label{xi1}
    \xi_1 =   6\gamma^3 \left({e \over m_0 c}\right)^2  \sum_{n,m,k} \Biggl[
        {| \{p_z z\}_{h1,ln}|^2\over (E_{ln}-E_{h1})^2}
        + {6\gamma\over m_0} {\{p^z z\}_{h1,ln}p^z_{ln,hm}p^z_{hm,lk}\{p^z z\}_{lk,h1} \over (E_{ln}-E_{h1})(E_{hm}-E_{h1})(E_{lk}-E_{h1})}
        \Biggr] \:,
\end{equation}
\begin{equation}
    \xi_2 =  {6\gamma^2 e^2 \over m_0 c^2} \sum_{m,n} \Biggl[
    {1\over 2}{I_{h1,ln}(z^2)_{ln,h1}\over E_{ln} - E_{h1}}
    - {\gamma_1-\gamma \over m_0} {p^z_{h1,ln}z_{ln,lm}\{p_z z\}_{lm,h1} \over (E_{ln}-E_{h1})(E_{lm}-E_{h1})}
    + {6\gamma^2 \over m_0^2} {p^z_{h1,ln}\{p^z z\}_{ln,h1}p^z_{h1,lm}\{p^z z\}_{lm,h1} \over (E_{ln}-E_{h1})(E_{hm}-E_{h1})(E_{lk}-E_{h1})}
    \Biggr] ,
\end{equation}
\end{widetext}
and $I_{h1,ln}=\langle h1 | ln \rangle$ are the overlap integrals of the
envelope functions.
We note that the dominant contribution to the cyclotron mass
renormalization comes from the coupling of the $h1$ and $h2$
subbands by the in-plane magnetic field
given by the first term in $\xi$, Eq.~\eqref{xi},
and is similar to that in
electron systems.~\cite{Ando} However, in contrast to the conduction
band where the effective mass is modified only for the direction
perpendicular to the magnetic field, in hole systems both components
of the effective mass tensor are renormalized.

Equations~\eqref{alpha_s} and~\eqref{xi} enable one to calculate the renormalizations of the cyclotron mass and Zeeman splitting in hole systems caused by the in-plane magnetic field.

\section{Discussion and Summary}\label{sec:discussion}

The experimental results discussed above demonstrate that, in
SiGe/Ge/SiGe quantum wells, the in-plane component of the magnetic
field leads to a decrease of the effective $g_z$ factor and to an
increase in the hole cyclotron mass. The effects are described by
Eqs.~\eqref{g_phen} and~\eqref{m_phen}, respectively, which also
follow from the microscopic theory, with the fitting parameters
${\alpha_s^{({\rm exp})} \approx 1.4 \times 10^{-3}}$~T$^{-2}$ and
${\alpha_c^{({\rm exp})} \approx 6 \times 10^{-3}}$~T$^{-2}$.

\begin{figure}[b]
\centerline{
\includegraphics[width=0.49\linewidth]{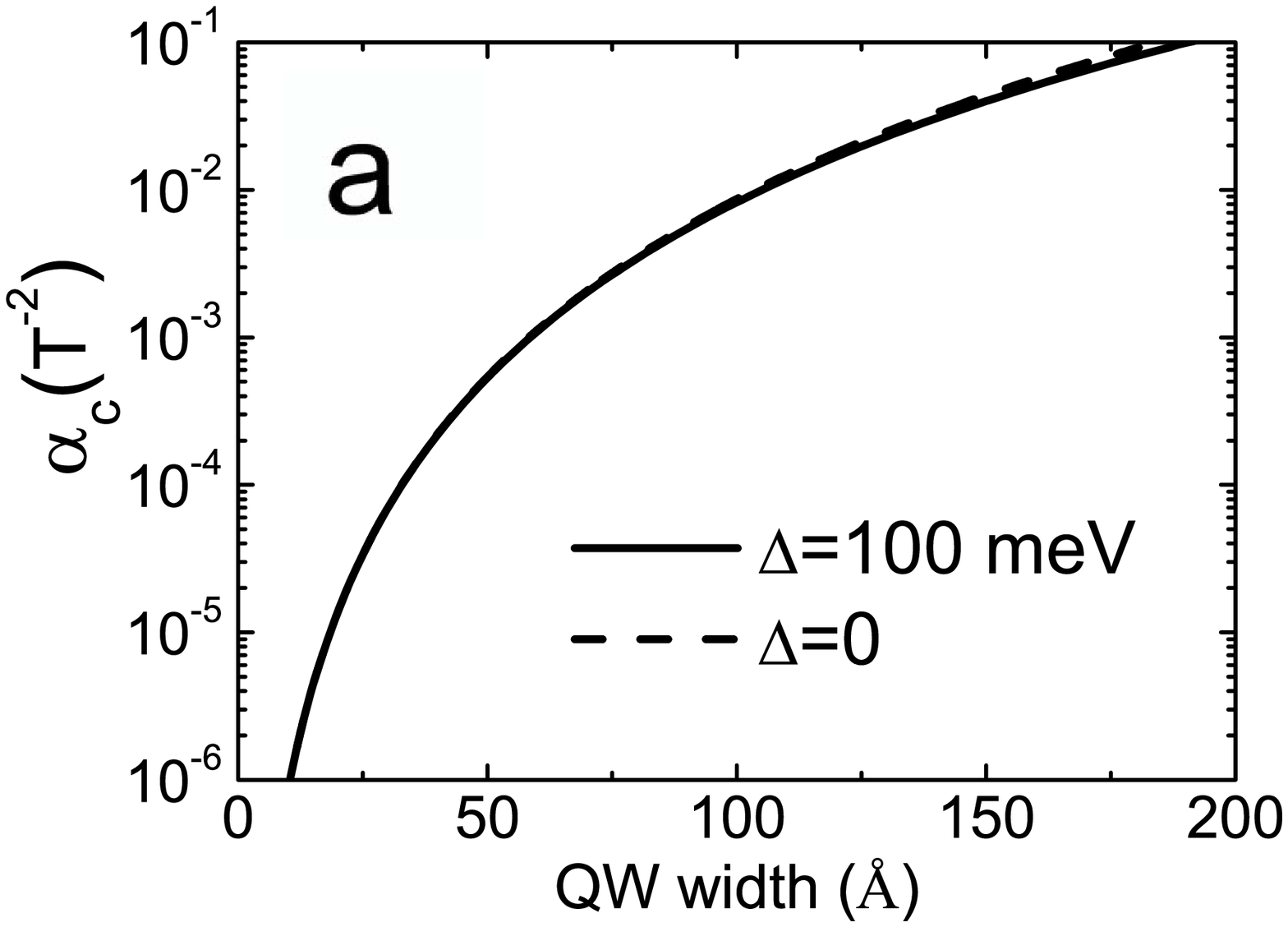}
\includegraphics[width=0.49\linewidth]{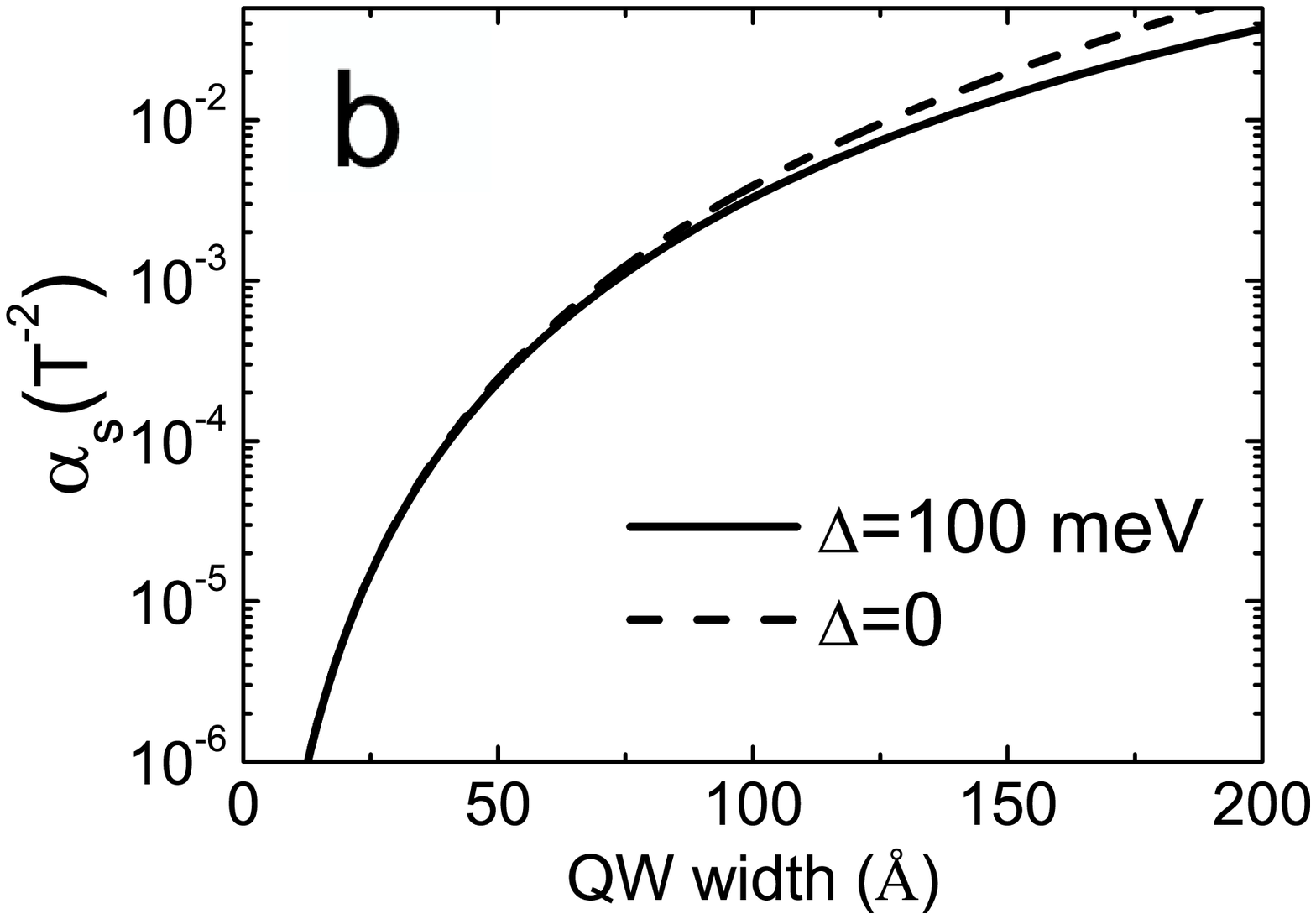}
} \caption {Dependence of (a) $\alpha_c$ and (b) $\alpha_s$
on the QW width for a strained ($\Delta=100$~meV, solid curves) and strain-free ($\Delta=0$, dashed curves) Ge quantum well. The curves are calculated for the Luttinger parameters $\gamma_1=13$ and $\gamma=5$.}
\label{fig_alpha_cs_vs_QWwidth}
\end{figure}

Figure~\ref{fig_alpha_cs_vs_QWwidth} shows the theoretical
dependence of the coefficients $\alpha_c$ and $\alpha_s$ describing
the effective mass and $g_z$ factor renormalization, respectively,
on the QW width. The curves are calculated for a rectangular Ge
quantum well with the infinitely high barriers, taking into account
the strain-induced splitting of the hh and lh subbands (solid
curves) and neglecting the strain (dashed curves). Both $\alpha_c$
and $\alpha_s$ increase with the QW width. The coefficient
$\alpha_c$ is almost independent of strain since its dominant part
is determined by the structure of the hh subbands only. In contrast,
the Zeeman splitting renormalization occurs due to the mixing of the
hh and lh subbands and $\alpha_s$ is therefore sensitive to the
hh-lh splitting induced by strain. The dependence of $\alpha_s$ on
strain is more pronounced for wide QWs where the hh-lh splitting due
to size quantization is comparable to or smaller than the
strain-induced splitting. The absolute values of $\alpha_c$ and
$\alpha_s$ extracted from the experiment correspond to the
calculated values for the QW width of about $100$~\AA~  according to
Fig.~\ref{fig_alpha_cs_vs_QWwidth}. The nominal QW width in the
studied sample is $200$~\AA, but the effective length of the hole
confinement might be considerably smaller due to the build-in
electric field produced by ionized dopants incorporated in the
barrier above the QW (see Sec.~\ref{Sample_Method}). This electric
field pushes the holes to the upper interface thereby reducing the
effective QW width.
%


To summarize, we have performed contactless acoustoelectric
measurements of the high-frequency conductivity of the
two-dimensional hole gas in a $p$-SiGe/Ge/SiGe structure subjected
to a strong magnetic field. It has been shown that in certain
temperature domains and integer filling factors the conductivity is
of the activation nature. The analysis of the activated conductivity
at odd and even filling factors in perpendicular magnetic field
enabled us to determine
$|g_z| \approx 6.7$. By
applying a tilted magnetic field
we observed that at fixed normal component of the field, the
conductivity in oscillation minima increases with increasing
$\Theta$ ( $\Theta  > 60^\circ$) at both even and odd filling
factors for $\nu \geq$ 5. Such a behavior is attributed to the
decrease of the cyclotron frequency and Zeeman splitting of holes
and the broadening of the Landau levels  by the in-plane component
of the magnetic field. We have developed a microscopic theory of the
heavy-hole cyclotron mass and $g_z$ factor renormalization in the
framework of the Luttinger Hamiltonian.  This theory describes the
experimental data and predicts the renormalization to be more
pronounced in wide quantum wells.

\acknowledgments
The authors would like to thank E. Palm, T. Murphy, J.-H. Park, and
G. Jones for technical assistance and V.A.~Volkov and V.S.~Khrapai
for useful discussions. This work was supported by Russian
Foundation for Basic Research,
the Program "Spintronika" of Branch of Physical Sciences of RAS,
and the U.M.N.I.K grant 16906. National High Magnetic Field
Laboratory is supported by NSF Cooperative Agreement No.
DMR-1157490, the State of Florida, and the U.S. Department of
Energy.


\begin{thebibliography}{99}
\bibitem{footnote1} The vanishingly small value of the
in-plane $g$-factor $g_\parallel$ is additionally supported by the
fact that no change in the ac conductivity is observed in the pure
in-plane magnetic field.
\end{thebibliography}
\end{document}